\documentclass{article}
\usepackage{amscd,amsmath,amssymb,bbm}
\usepackage{amsfonts}
\newcommand{\p}[1]{(\ref{#1})}

\newcommand{\bD}{{\overline{\strut D}}{}}

\newcommand{\tb}{{\bar\theta}}

\newcommand{\be}{\begin{equation}}
\newcommand{\ee}{\end{equation}}
\newcommand{\bea}{\begin{eqnarray}}
\newcommand{\eea}{\end{eqnarray}}
\newcommand{\ba}{\begin{array}}
\newcommand{\ea}{\end{array}}

\newcommand{\und}{\qquad\textrm{and}\qquad}

\newcommand{\nn}{\nonumber}

\newcommand{\R}{{\mathbb{R}}}

\def\sfrac#1#2{{\textstyle\frac{#1}{#2}}}

\def\di{{\rm d}}
\def\im{{\rm i}}
\def\ep{{\rm e}}
\def\={\ =\ }

\def\Nf{$\cal N${=\,}4~}
\def\Ne{$\cal N${=\,}8~}

\begin{document}
\thispagestyle{empty}
\vspace{2cm}
\begin{flushright}
ITP--UH--09/09\\
\end{flushright}\vspace{2cm}
\begin{center}
{\LARGE\bf SU(2) reduction in \Nf supersymmetric mechanics}
\end{center}
\vspace{1cm}

\begin{center}
{\Large
Sergey Krivonos$\,{}^{a}$ \ \ and \ \
Olaf Lechtenfeld$\,{}^{b}$
}
\vspace{1.0cm}

${}^a$
{\it Bogoliubov  Laboratory of Theoretical Physics,
JINR, 141980 Dubna, Russia}
\vspace{0.2cm}

${}^b$
{\it  Institut f\"ur Theoretische Physik, Leibniz Universit\"at Hannover,
30167 Hannover, Germany}
\vspace{0.5cm}

{\tt krivonos@theor.jinr.ru, lechtenf@itp.uni-hannover.de}
\end{center}
\vspace{3cm}

\begin{abstract}
\noindent
We perform an $su(2)$ Hamiltonian reduction of the general $su(2)$-invariant
action for a self-coupled $(4,4,0)$ supermultiplet. As a result,
we elegantly recover the \Nf supersymmetric mechanics with spin degrees of
freedom which was recently constructed in {\tt arXiv:0812.4276}. 
This observation underscores the exceptional role played by 
the ``root'' supermultiplet in \Nf supersymmetric mechanics.
\end{abstract}

\newpage
\setcounter{page}{1}

\setcounter{equation}0
\section{Introduction}
In a recent paper \cite{fil0}, \Nf superconformal mechanics with $n$ bosonic
and $4n$ fermionic degrees of freedom has been endowed with a potential term
through a coupling to auxiliary supermultiplets with $4n$ bosonic and
$4n$ fermionic components~\cite{DI}. This combination gave rise to an
OSp($4|2$) supersymmetric $n$-particle Calogero model. Subsequently, the
one-particle case, i.e.~OSp($4|2$) superconformal mechanics, was analyzed
on the classical and quantum level~\cite{fil}. Simultaneously, it was
demonstrated that the potential-generating strategy works perfectly for the
most general $D(2,1;\alpha)$ superconformal one-particle mechanics~\cite{bk}.
It is quite satisfying how the spin degrees of freedom appear in the bosonic
sector, with only first time derivatives in the action. Thus, the proposed
coupling of two different \Nf supermultiplets provides a simple and elegant
way to incorporate spin degrees of freedom in supersymmetric mechanics.

In both previous treatments~\cite{fil,bk}, on mass-shell all components of
the basic $(1,4,3)$ supermultiplet are expressed through those of the
``auxiliary'' $(4,4,0)$ one. It seems that just this ``auxiliary''
supermultiplet plays a fundamental role in the construction.
It is therefore natural to inquire whether the these models can be reformulated
{\it purely\/} in terms of $(4,4,0)$ supermultiplets. Of course,
such a reformulation has to be supplied with a Hamiltonian reduction,
which would reduce the four physical bosons to one boson plus spin variables.
Alternatively, the passage from SU(2)-symmetric $(4,4,0)$ models to general 
$(1,4,3)$ models via gauging was described in~\cite{DI} using harmonic
superspace.

Incidentally, spin degrees of freedom have appeared in a bosonic system
after Hamiltonian reduction (on the Lagrangian level) via the second Hopf
map $S^7/S^3\simeq S^4$~\cite{armen}.
In the bosonic sector this reduced system resembles those in~\cite{fil,bk},
besides the presence of four additional bosonic variables.

In the present note we realize the above ideas and rederive
the \Nf supersymmetric ``spin mechanics'' of~\cite{fil,bk} by an $su(2)$
Hamiltonian reduction applied to the general $su(2)$ invariant action
for a self-coupled (4,4,0) supermultiplet. It is a further manifestation
of the fundamental importance of the ``root'' supermultiplet~\cite{FG} in
\Nf supersymmetric mechanics \cite{root,DI,DI1}.

\setcounter{equation}0
\section{SU(2) reduction}
Our point of departure is a quartet of real \Nf superfields
$Q^{ia}$ with $i,a=1,2$ defined in the \Nf superspace
$\R^{(1|4)}=(t,\theta_i,\tb{}^i)$ and subject to the constraints
\be\label{con}
D^{(i}Q^{j)a}=0\ , \qquad \bD^{(i}Q^{j)a}=0 \und
\left( Q^{ia}\right)^\dagger = Q_{ia}\ ,
\ee
where the corresponding covariant derivatives have the form
\be
D^i=\frac{\partial}{\partial \theta_i}+\im\tb{}^i \partial_t\ ,\qquad
\bD_i=\frac{\partial}{\partial \tb{}^i}+\im\theta_i \partial_t
\qquad\textrm{so that}\qquad
\bigl\{ D^i, \bD_j\bigr\}=2\im \delta^i_j\partial_t\ .
\ee
This \Nf supermultiplet describes four bosonic and four fermionic but zero
auxiliary variables off-shell~\cite{hyper, ikl1}.
Let us now introduce the composite \Nf superfield
\be\label{X}
X\=2\,(Q^{ia}Q_{ia})^{-1}
\ee
which, in virtue of \p{con}, obeys the constraints~\cite{ikl1}
\be\label{con1}
D^i D_i X\= \bD_i \bD{}^i X \= \bigl[ D^i, \bD_i \bigr] X\=0\ .
\ee
The most general action for $Q^{ia}$ is constructed by integrating an arbitrary
superfunction $\widetilde{\cal F}(Q^{ia})$ over the whole \Nf superspace.
Here, we restrict ourselves to prepotentials of the form
\be\label{action1}
\widetilde{\cal F}(Q^{ia})\={\cal F}\bigl(X(Q^{ia})\bigr)
\qquad\longrightarrow\qquad
S\=-\sfrac{1}{8}\int\!\!\di{t}\,\di^4\theta\ {\cal F}(X)\ .
\ee
The rationale for this selection is its manifest invariance under $su(2)$
transformations acting on the ``$a$'' index of~$Q^{ia}$. This is the symmetry
over which we are going to perform the Hamiltonian reduction.

In terms of components the action \p{action1} reads
\be\label{action2}
S\=\int\!\!\di{t}\ \Bigl\{ G \bigl( {\dot x}{}^2 +
\im({\dot\eta}{}^i\bar\eta_i-\eta^i\dot{\bar\eta}_i)+
\sfrac{1}{2}x^2 \omega^{ij}\omega_{ij}\bigr) -
\im\bigl(2G+xG'\bigr)\omega^{ij}\eta_i\bar\eta_j -
\sfrac{1}{4}\bigl(G''+6\sfrac{G'}{x}+6\sfrac{G}{x^2}\bigr)
\eta^i\eta_i\bar\eta_j\bar\eta{}^j\Bigr\}\ ,
\ee
where
\be\label{comp}
x=X|\ ,\qquad \eta^i=-\im D^i X|\ ,\qquad
\bar\eta_i =-\im\bD_i X|\ ,\qquad q^{ia}=\sqrt{X} Q^{ia}|\ ,\qquad
G={\cal F}{}''(X)|
\ee
and
\be\label{def}
\omega_{ij}\={\dot q}{}^a_i q_{ja}+{\dot q}^a_j q_{ia} \ .
\ee
Here, as usually, $(\ldots)|$ denotes the $\theta_i=\bar\theta_i=0$ limit.

To proceed we introduce the following substitution
for the  bosonic variables $q^{ia}$ subject to $q^{ia}q_{ia}=2$,
\be\label{q}
q^{11}= \frac{\ep^{-\frac{\im}2\phi}}{\sqrt{1{+}\Lambda\bar\Lambda}}\,\Lambda
\ ,\qquad
q^{21}=-\frac{\ep^{-\frac{\im}2\phi}}{\sqrt{1{+}\Lambda\bar\Lambda}}
\ , \qquad
q^{22}= \left(q^{11}\right)^\dagger\ ,\qquad
q^{12}=-\left(q^{12}\right)^\dagger\ .
\ee
In terms of the new variables $(\phi,\Lambda,\bar\Lambda)$ the $su(2)$ 
rotations \ $\delta q^{ia}=\gamma^{(ab)}q^i_b$ \ read~\cite{ikl1}
\be\label{transf1}
\delta\Lambda = \gamma^{11}\ep^{\im\phi}(1{+}\Lambda\bar\Lambda)\ ,\qquad
\delta\bar\Lambda = \gamma^{22}\ep^{-\im\phi}(1{+}\Lambda\bar\Lambda)\ ,\qquad
\delta\phi=-2\im\gamma^{12}+
\im\gamma^{22}\ep^{-\im\phi}\Lambda -\im\gamma^{11}\ep^{\im\phi}\bar\Lambda\ .
\ee
It is easy to check that
\be\label{omega}
\omega^{11}\=2\frac{\dot\Lambda-\im\Lambda\dot\phi}{1+\Lambda\bar\Lambda}
\ ,\qquad
\omega^{22}\=\left( \omega^{11}\right)^\dagger \und
\omega^{12}\=\im\frac{1-\Lambda\bar\Lambda}{1+\Lambda\bar\Lambda}\,\dot\phi\ +\
\frac{\dot\Lambda\bar\Lambda-\Lambda\dot{\bar\Lambda}}{1+\Lambda\bar\Lambda}
\ee
are indeed invariant under~\p{transf1}, as is the whole action \p{action2}.

Next, we introduce the standard Poisson brackets
\be\label{pb}
\left\{\pi,\Lambda\right\}=1\ ,\qquad
\left\{\bar\pi,\bar\Lambda\right\}=1\ ,\qquad
\left\{p_{\phi},\phi\right\}=1\ ,
\ee
so that the generators of the transformations \p{transf1},
\be\label{currents}
I_{\phi}\=p_{\phi}\ ,\qquad
I\=\ep^{\im\phi}\bigl[(1{+}\Lambda\bar\Lambda)\,\pi-
\im\bar\Lambda\,p_\phi\bigr] \ ,\qquad
{\bar I}\=\ep^{-\im\phi}\bigl[(1{+}\Lambda\bar\Lambda)\,\bar\pi+
\im\Lambda\,p_\phi\bigr]\ ,
\ee
will be the Noether constants of motion for the action~\p{action2}.
To perform the reduction over this SU(2) group we fix the Noether constants
as~(c.f.~\cite{armen})
\be\label{reduction1}
I_\phi=m \und  I={\bar I}=0\ ,
\ee
which yields
\be\label{reduction2}
p_\phi\=m \und
\pi\=\frac{\im m\,\bar\Lambda}{1+\Lambda\bar\Lambda}\ ,\qquad
\bar\pi\=-\frac{\im m\,\Lambda}{1+\Lambda\bar\Lambda}\ .
\ee
Conducting a Routh transformation over the variables
$(\Lambda,\bar\Lambda,\phi)$, we reduce the action~\p{action2} to
\be\label{action3}
{\widetilde S}\ \=\ S\ -\ \int\!\!\di{t}\ \bigl\{
\pi\,{\dot\Lambda}+\bar\pi\,\dot{\bar\Lambda}+p_\phi{\dot\phi}\bigr\}
\ee
and substitute the expressions~\p{reduction2} into~$\tilde S$.
A slightly lengthy but straightforward calculation gives
\bea\label{action4}
{\widetilde S}_{\rm red}&{=}& \int\!\!\di{t}\ \Bigl\{
G\bigl({\dot x}{}^2 +\im({\dot\eta}{}^i\bar\eta_i-\eta^i\dot{\bar\eta}_i)\bigr)
-\sfrac{1}{4}\bigl( G''-\sfrac{3}{2}\sfrac{(G')^2}{G}\bigr)\eta^2\bar\eta{}^2
-\sfrac{m^2}{4x^2 G} \Bigr. \nn\\
&&\qquad\Bigl.-\ \sfrac{m(2G+x G')}{2x^2 G(1{+}\Lambda\bar\Lambda)}
\bigl(2\Lambda\eta_1\bar\eta_1-2\bar\Lambda\eta_2\bar\eta_2-
(1{-}\Lambda\bar\Lambda)(\eta_1\bar\eta_2+\eta_2\bar\eta_1)\bigr)\Bigr\}\ .
\eea
To ensure that the reduction constraints \p{reduction2} are satisfied
we add Lagrange multiplier terms,
\be\label{action5}
S_{\rm red}\ \=\ {\tilde S}_{\rm red}\ +\ \int\!\!\di{t}\ \Bigl\{m\,\dot\phi
\ +\ \sfrac{\im m\,(\dot\Lambda\bar\Lambda-\Lambda\dot{\bar\Lambda})}
{1+\Lambda\bar\Lambda}\Bigr\}\ .
\ee
Finally, by employing new variables $v^i=q^{i1}$ and ${\bar v}_i=(v^i)^\dagger$
we rewrite this action in the symmetric form
\bea\label{action6}
S_{\rm red}&{=}& \int\!\!\di{t}\ \Bigl\{
G\bigl({\dot x}{}^2 +\im({\dot\eta}{}^i\bar\eta_i-\eta^i\dot{\bar\eta}_i)\bigr)
-\sfrac{1}{4}\bigl( G''-\sfrac{3}{2}\sfrac{(G')^2}{G}\bigr)\eta^2\bar\eta{}^2
-\sfrac{m^2}{4x^2 G} \Bigr. \nn\\
&&\qquad\Bigl.+\ \im m\,({\dot v}{}^i{\bar v}_i-v^i\dot{\bar v}_i)
-\sfrac{m(2G+x G')}{2x^2 G }v^i{\bar v}{}^j (\eta_i\bar\eta_j+\eta_j\bar\eta_i)
\Bigr\} \qquad\textrm{with}\quad v^i{\bar v}_i=1\ .
\eea
Amazingly, this final action coincides with the one presented in~\cite{bk}
and specializes to the one derived in~\cite{fil} for the choice of $G=1$,
which corresponds to OSp($4|2$) symmetry.

We stress that the $su(2)$ reduction algebra, realized in~\p{transf1},
commutes with all (super)symmetries of the action~\p{action1}.
Therefore, all symmetry properties of the theory
(including the $D(2,1;\alpha)$ invariance for a properly chosen
prepotential ${\cal F}$) are preserved in our reduction.

\section{Conclusion}
We have demonstrated that the novel \Nf supersymmetric ``spin mechanics''
of~\cite{fil0,fil,bk} is nicely interpreted as an $su(2)$ reduction of a
self-interacting ``root'' supermultiplet with $(4,4,0)$ component content.
This procedure is remarkably simple and automatically successful.

An almost straightforward application of this insight is a similar $su(2)$
reduction applied to the \Nf ``nonlinear''~supermultiplet \cite{ikl1}.
The resulting system will contain only spinor variables accompanied by four
fermions. In this regard, one could also investigate the nonlinear ``root''
supermultiplet and its action~\cite{bkls}.

Finally, we mention that our reduction will almost never work for the
\Ne supersymmetric mechanics in the literature.
The reason is simple: these systems do not possess any internal symmetry
which commutes with all eight supersymmetries. This is also the situation
discussed in~\cite{armen}. The one positive exception is the ``real''
${\cal N}{=\,}8$, $d{=}1$ hypermultiplet, which is obtained by dimensional 
reduction from ${\cal N}{=\,}2$, $d{=}4$ and requires ${\cal N}{=\,}8$, $d{=}1$
harmonic superspace~\cite{HSS,HSS1}.
We expect the corresponding $su(2)$ reduction to produce some spin extension
of the recently constructed \Ne superconformal mechanics~\cite{DI22}.
We intend to turn to this issue soon.

\section*{Acknowledgements}
We thank Evgeny Ivanov for comments.
S.K.\ is grateful to the ITP at Leibniz Universit\"at Hannover for hospitality.
This work was partially supported by the grants
RFBR-08-02-90490-Ukr, 06-02-16684, and DFG grant~436 Rus~113/669/03.


\bigskip

\begin{thebibliography}{99}
\bibitem{fil0}
S.~Fedoruk, E.~Ivanov, O.~Lechtenfeld,\\
{\it Supersymmetric Calogero models by gauging,}
{\tt arXiv:0812.4276 [hep-th]}.
\bibitem{DI}
F.~Delduc, E.~Ivanov,
Nucl. Phys. B770 (2007) 179 {\tt arXiv:hep-th/0611247}.
\bibitem{fil}
S.~Fedoruk, E.~Ivanov, O.~Lechtenfeld,\\
{\it OSp(4$|$2) Superconformal Mechanics,}
{\tt arXiv:0905.4951 [hep-th]}.
\bibitem{bk}
S.~Bellucci, S.~Krivonos,
{\it Potentials in N=4 superconformal mechanics,}
{\tt arXiv:0905.4633 [hep-th]}.
\bibitem{armen}
M.~Gonzales, Z.~Kuznetsova, A.~Nersessian, F.~Toppan, V.~Yeghikyan,\\
{\it Second Hopf map and supersymmetric mechanics with Yang monopole,}
{\tt arXiv:0902.2682 [hep-th]}.
\bibitem{FG}
M.~Faux, S.J.~Gates Jr.,
Phys. Rev. D71 (2005) 065002, {\tt arXiv:hep-th/0408004}.
\bibitem{root}
S.~Bellucci, S.~Krivonos, A.~Marrani, E.~Orazi,\\
Phys. Rev. D73 (2006) 025011, {\tt arXiv:hep-th/0511249}.
\bibitem{DI1}
F.~Delduc, E.~Ivanov,
Nucl. Phys. B753 (2006) 211, {\tt  arXiv:hep-th/0605211};\\
Nucl. Phys. B787 (2007) 176, {\tt  arXiv:0706.0706 [hep-th]}.
\bibitem{hyper}
E.~Ivanov, O.~Lechtenfeld,
JHEP 0309 (2003) 073, {\tt arXiv:hep-th/030711}.
\bibitem{ikl1}
E.~Ivanov, S.~Krivonos, O.~Lechtenfeld,
Class. Quant. Grav. 21 (2004) 1031, {\tt arXiv:hep-th/0310299}.
\bibitem{bkls}
S.~Bellucci, S.~Krivonos, O.~Lechtenfeld, A.~Shcherbakov,\\
Phys. Rev. D77(2008)045026, {\tt arXiv:0710.3832 [hep-th]}.
\bibitem{HSS}
A.S.~Galperin, E.A.~Ivanov, V.I.~Ogievetsky, E.S.~Sokatchev,\\
{\it Harmonic superspace}, Cambridge University Press, 2001.
\bibitem{HSS1}
E.~Ivanov, O.~Lechtenfeld,
JHEP 0309 (2003) 073, {\tt arXiv:hep-th/0307111}.
\bibitem{DI22}
F.~Delduc, E.~Ivanov,
Phys. Lett. B654 (2007) 200, {\tt arXiv:0706.2472 [hep-th]}.

\end{thebibliography}
\end{document}